\documentclass{cpp2e}
\usepackage{epsfig, floatflt, rotating, latexsym, amssymb}
\shorttitle{Dielectric properties of correlated quantum plasmas} 
\shortauthor{M.~Bonitz et al.} 
\contvolum{41} 
\contyear{2001} 
\contnumber{2/3} 
\contpages{155-158} 
\title{
Dielectric properties of correlated quantum plasmas
}
\author{M.~Bonitz$^1$,
V.~Golubnichiy$^1$, N.H.~Kwong$^2$, D.~Semkat$^1$, D.~Kremp$^1$,
V.S.~Filinov$^3$, and M.~Schlanges$^4$ }
\address{
$^1$Fachbereich Physik, Universit{\"a}t Rostock,
Universit{\"a}tsplatz 3, D-18051 Rostock \\
$^2$ Optical Sciences Center, University of Arizona, Tucson, AZ 85721, USA \\
$^3$ High Energy Density Research Center Rus.Ac.Sc.,
Izhorskaya street 13-19, Moscow 127412, Russia \\
$^4$Institut f\"ur Physik der Universit{\"a}t Greifswald,
Domstrasse 10a, D-17489 Greifswald  \\
e-mail:michael.bonitz@physik.uni-rostock.de}
\begin{document}
\setcounter{page}{155}
\makeheadings
\maketitle
\begin{abstract}
Results for the dynamic structure factor of a one-component plasma
are presented for the situations of strong coupling and weak degeneracy
and strong degeneracy and weak coupling. Possibilities to obtain
rigorous results when both, coupling and degeneracy are large are
discussed. \\
\end{abstract}


In dense astrophysical and laboratory plasmas one frequently encounters
situations where Coulomb and quantum effects are important simultaneously \cite{sccs},
i.e. where the Coulomb coupling parameter
$\Gamma=(4\pi n_e/3)^{1/3}e^2/4\pi \epsilon_0 k_BT$ and the degeneracy parameter
$\chi=n_e\lambda_e^3$ [$\lambda_e$ is the electron
thermal wave length $\lambda_e^2=2\pi\hbar^2 \beta/m_e$] exceed unity. While
a rigorous theoretical treatment of the equilibrium properties is now becoming
possible, in particular by path integral quantum Monte Carlo (PIMC) techniques
\cite{fb}, comparable results for the {\em dynamic properties} are still missing.
In this paper, we discuss two possible approaches to solve this problem. We
focus on the dynamic dielectric function $\epsilon(\omega,q)$ and the dynamic
structure factor $S(\omega,q)\sim -{\rm Im}\,1/\epsilon(\omega,q)$ from which all
optical and transport properties can be computed.

The central quantity which describes dynamical properties of an N-particle system
is the {\em density fluctuation} $\rho_{\bf q}$, where ${\bf q}$ is the wave vector.
In a classical system, it
can be computed  from Klimontovich's phase space density
${\hat \rho}_{{\bf p},{\bf r}}(t)=
   \sum_{i=1}^N
   \delta[{\bf r}-{\bf {\hat r}}_i(t)]\,
   \delta[{\bf p}-{\bf {\hat p}}_i(t)],
$ where ${\hat {\bf r}}_i(t)$ and ${\hat {\bf p}}_i(t)$ are classical (random) particle
trajectories. Fourier transformation with respect to ${\bf r}$ yields the momentum-dependent
density fluctuation $\rho_{\bf p,\bf q}$, and the total fluctuation
$\rho_{\bf q}$ is obtained by  a subsequent integration over all momenta,
\begin{eqnarray}
{\hat \rho}_{{\bf p},{\bf q}}(t)=
   \sum\limits_{i=1}^N
   e^{i{\bf q}{\bf {\hat r}}_i(t)}\,
   \delta[{\bf p}-{\bf {\hat p}}_i(t)], \qquad
{\hat \rho}_{{\bf q}}(t)=
   \sum\limits_{i=1}^N
   e^{i{\bf q}{\bf {\hat r}}_i(t)}.
\label{rho_c}
\end{eqnarray}

The corresponding quantum result is derived from the N-particle wave function
${\hat \Psi}_{N}({\bf r},t)=\sum_{s} {\hat a}_s(t)\phi^{(s)}_N({\bf r},t)$, where
$\phi^{(s)}_N$ is a complete set of basis functions, and the coefficients are
creation/annihiliation operators obeing
(anti-)commutation rules $[{\hat a}^{\dagger}_{s}(t),{\hat a}_{s'}(t)]_{\mp}=\delta_{s,s'}$
for bosons (fermions). The quantum generalization of Eq.~({\ref{rho_c}) is then given by
($\hbar=1$)
\begin{eqnarray}
{\hat \rho}_{{\bf p},{\bf q}}(t)=
{\hat a}^{\dagger}_{\bf p}(t_2)\,{\hat a}_{{\bf p}+{\bf q}}(t_1)\big |_{t_1=t_2=t}, \qquad
{\hat \rho}_{{\bf q}}(t)=
\sum\limits_{{\bf p}}\, {\hat a}^{\dagger}_{\bf p}(t)\,{\hat a}_{{\bf p}+{\bf q}}(t),
\label{rho_q}
\end{eqnarray}
where the fluctuation ${\hat \rho}_{{\bf p},{\bf q}}$ describes the annihilation of an
electron in momentum state ${\bf p}+{\bf q}$ at time $t_1$ and creation
in momentum state ${\bf p}$ at time $t_2$. Obviously, the limiting case $q=0$ of
Eqs.~(\ref{rho_c}) and (\ref{rho_q}) yields the momentum distribution and the
density.

The fluctuations ${\hat \rho}_{{\bf p},{\bf q}}$ and ${\hat \rho}_{{\bf q}}$ are random
quantities depending on the initial conditions of the equations of motion. To come to
measurable quantities, a suitable {\em average} over the particle ensemble (e.g. over the
initial conditions) has to be performed yielding
$\rho_{{\bf p},{\bf q}}(t)\equiv\langle {\hat \rho}_{{\bf p},{\bf q}}(t)\rangle$ and
$\rho_{{\bf q}}(t)\equiv\langle {\hat \rho}_{{\bf q}}(t)\rangle$. These are already
macroscopic observables which are closely related to the dynamic structure factor.
There exist two ways to compute $S(\omega,q)$ for a N-particle system in {\em equilibrium}.

The {\bf first} is to analyze the {\bf equilibrium fluctuations in a field-free system}.
In this case, $\rho_{{\bf q}}(t)\approx 0$ due to spatial homogeneity. The first non-vanishing
average is the density-density correlation function $C_{\rho\rho}$ which is closely
related to pair correlations and the Fourier transform of which yields the dynamic
structure factor,
\begin{eqnarray}
C_{\rho\rho}({\bf q},\tau) \equiv
\langle {\hat \rho}_{{\bf q}}(\tau){\hat \rho}_{-{\bf q}}(0)\rangle, \qquad
S(\omega,{\bf q}) = \frac{1}{2\pi}\int_{-\infty}^{\infty} d\tau \,e^{i\omega \tau} \,
C_{\rho \rho}({\bf q},\tau).
\label{s_c}
\end{eqnarray}
$C_{\rho\rho}$ can be computed from equilibrium theories, such as the Bethe Salpeter
equation, or nonequilibrium approaches, in particular molecular dynamics simulations
(MD). The latter has the advantage that, for a classical system, {\em correlations of arbitrary
strength} are straightforwardly included, e.g. in the frame of Newton's equations. After
averaging the resulting trajectories ${\hat {\bf r}}_i(t)$ and ${\hat {\bf p}}_i(t)$ [over
runs with different initial conditions or over time], the dynamic structure factor follows
directly from Eqs.~(\ref{rho_c}) and (\ref{s_c}).

We have applied this approach to a degenerate electron gas by using an effective
quantum pair potential in the classical MD scheme. An appropriate potential was obtained by
Kelbg \cite{kelbg} and is rigorous for $\Gamma<1$ and $\chi<1$. We have performed extensive
simulations in which the influence of quantum effects on the dynamic structure factor
and the plasmon spectrum has been investigated \cite{md_ocp}. It is found that quantum
effects influence the plasmon spectrum predominantly at large wave numbers, but their
influence is rather small. Fig.~1.a shows an example of the simulation results and a
comparison to mean-field theories.

The {\bf second approach} to dynamic properties is to investigate the
{\bf temporal response of the
plasma to a weak inhomogeneous field} $U(t)$ with wave number $q_o$. The field gives rise to
collective oscillations of the electrons and thus to a non-vanishing fluctuation
$\rho_{{\bf q}_o}(t)$ the Fourier transform of which directly yields the dynamic
structure factor \cite{kwong-etal.00prl}
\begin{eqnarray}
S(\omega,{\bf q}_o) \sim
- \frac{1}{n}\,\frac{\rho_{{\bf q}_{o}}(\omega)}{U_{{\bf q}_o}(\omega)}.
\label{s_q}
\end{eqnarray}
This expression is much simpler than (\ref{s_c}) and well suited for classical MD simulations.
On the other hand, using the quantum expressions (\ref{rho_q}), a kinetic theory of the
dynamic properties of {\em strongly degenerate systems} can be developed. Starting from the
Heisenberg equation of motion
$
i \,\frac{\partial}{\partial t}{\hat \rho}_{{\bf p},{\bf q}} -
[{\hat \rho}_{{\bf p},{\bf q}}, {\hat H}^{\rm sys}+{\hat H}^{\rm ext}]=0,
$
with the system and external hamiltonians
$
{\hat H}^{\rm sys} =
\sum_{\bf k}\epsilon_{\bf k}{\hat a}^{\dagger}_{\bf k}\,{\hat a}_{{\bf k}} +
{\normalsize\frac{1}{2}} \sum_{{\bf k}_1{\bf k}_2{\bf q}\ne 0}V(q)
  {\hat a}^{\dagger}_{{\bf k}_1+{\bf q}}{\hat a}^{\dagger}_{{\bf k}_2-{\bf q}}
  {\hat a}_{{\bf k}_2}{\hat a}_{{\bf k}_1},
$
$
{\hat H}^{\rm ext}(t)=
\sum_{\bf q} U({\bf q},t) \sum_{\bf k} {\hat a}^{\dagger}_{\bf k}\,{\hat a}_{{\bf k-q}},
$
one can derive an equation of motion for the
averaged fluctuation of $\rho_{{\bf p},{\bf q}}$. However, the most general theory is obtained
if one considers the {\em two-time fluctuations}, cf. Eq.~(\ref{rho_q}), [there are 2
independent functions due the non-commutivity of the operators $a, a^{\dagger}$]
$
G^<_{{\bf p}_1,{\bf p}_2}(t_1 t_2) \equiv
i \langle {\hat a}^{\dagger}_{{\bf p}_2}(t_2) {\hat a}_{{\bf p}_1}(t_1)  \rangle
$
and
$
G^>_{{\bf p}_1, {\bf p}_2}(t_1 t_2) \equiv
-i \langle {\hat a}_{{\bf p}_1}(t_1) {\hat a}^{\dagger}_{{\bf p}_2}(t_2) \rangle
$. The equations of motion for $G^{\gtrless}$ are the inhomogeneous
Kadanoff-Baym equations (KBE) and read
\begin{eqnarray}
\left( i \frac{\partial}{\partial t_1} -
\epsilon_{{\bf k}_1} \right)G_{{\bf k}_1,{\bf k}_2}^{\gtrless}(t_1 t_2)
=
\sum_{{\bf q}} U(-{\bf q},t_1)
G_{{\bf k}_1-{\bf q},{\bf k}_2}^{\gtrless}(t_1 t_2) +
\nonumber\\
\sum_{{\bar {\bf k}}} \Sigma_{{\bf k}_1,{\bar {\bf k}}}^{HF}(t_1 t_1)
G_{{\bar {\bf k}},{\bf k}_2}^{\gtrless}(t_1 t_2) +
I_{{\bf k}_1,{\bf k}_2}^{\gtrless}(t_1 t_2),
\label{kbeq}
\end{eqnarray}
and are to be supplemented with the adjoint equation. The Coulomb potential $V$ gives rise to
mean-field and short-range correlation effects which are contained in the
Hartree--Fock selfenergy $\Sigma^{HF}$ and the collision integrals $I$, respectively.

\begin{floatingfigure}{10.2cm}
	\epsfig{file=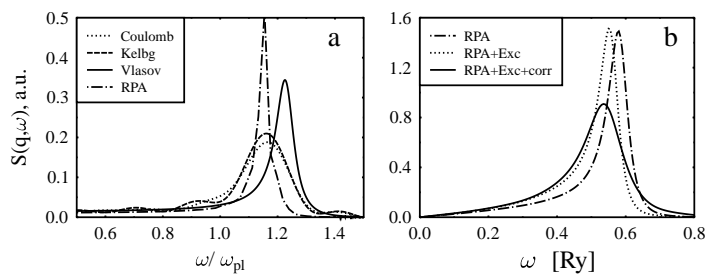, width=10cm}
\vspace{-1.cm}
	\caption{\small{Dynamic structure factor for an electron gas.
{\bf a:}
Using Eq.(3) with molecular dynamics simulations with the Coulomb and Kelbg
potential compared to the RPA and Vlasov theories for weak degeneracy,
$n\lambda^3=0.5$, $\Gamma=1$, $q{\bar r}=0.6$. {\bf b:} Quantum kinetic theory
results for $S$, Eq.(4), comparing RPA, RPA plus exchange and RPA with exchange
and collisions in Born approximation for strong degeneracy,
$r_s=4, T=0.16 E_R$, $q=0.6k_F$. ${\bar r}, r_s, E_R$ and $k_F$ denote the 
mean interparticld distance, Brueckner parameter, 
Hydrogen binding energy and Fermi momentum, respectively.}}
\label{fig1}
\end{floatingfigure}

Eqs.~(\ref{kbeq}) are very general and include the response from a nonequilibrium
state as well as the nonlinear response in case of a strong external field. But here
our interest is in the linear response to a weak field resulting in Eq.~(\ref{s_q}).
This limit is obtained by considering Eqs.~(\ref{kbeq}) for a monochromatic
excitation $U({\bf q},t)=U_0(t)\delta_{{\bf q},{\bf q}_o}$ of small amplitude $U_0$
which allows for a linearization. The result are coupled equations for the
spatially homogeneous and monochromatic components
$G^{\gtrless}_{00}\sim G_{{\bf k},{\bf k}}^{\gtrless}(t_1 t_2)$ and
$G^{\gtrless}_{10}\sim G_{{\bf k+q}_o,{\bf k}}^{\gtrless}(t_1 t_2)$, respectively,
where $|G^{\gtrless}_{00}|\sim U_0^0 \gg |G^{\gtrless}_{10}|\sim U_0^1$, for
details, cf. Ref. \cite{kwong-etal.00prl}. After
solving this system, the total fluctuation follows according to
$\rho_{{\bf q}_o}(t)=-i\sum_p G_{{\bf p+q}_o,{\bf p}}^<(t,t)$, the Fourier
transform of which yields the structure factor (\ref{s_q}). This is illustrated
in Fig. 2, middle and lower figures.

Equations (\ref{kbeq}) contain all important limiting cases of a quantum
electron plasma: i) neglect of
$\Sigma^{HF}$ and $I$ yields just the spectrum of single-particle excitations;
ii) inclusion of $\Sigma^H$ (Hartree mean field) is equivalent to the RPA
(or - in the classical limit - the Vlasov response); iii) inclusion of
$\Sigma^{HF}=\Sigma^H+\Sigma^F$ yields the RPA corrected by exchange
(time-dependent Hartree-Fock) and, finally, iv) addition of $I^{\gtrless}$
yields RPA with exchange and correlations. Numerical results comparing cases
ii)--iv) are shown in Fig.1.b.

Most remarkably, consistency
problems (e.g. sum rule preservation) often encountered in theories
incorporating collisions
in the dielectric response functions, are completely avoided due to the
conservation properties of the Kadanoff-Baym equations. In fact, choice of
density conserving collision integrals is sufficient to preserve the f-sum
rule. Moreover, the KBE easily allow to fully conserve the total energy,
i.e. all functions $G^{\gtrless}_{00},G^{\gtrless}_{10}$ are {\em full}
renormalized Greens functions and are thus appropriate to describe the
response from a correlated equilibrium or nonequilibrium state of the plasma.
Interestingly, the solution of the KBE in the time domain allows to account
\begin{floatingfigure}{8.cm}
	\mbox{\epsfig{file=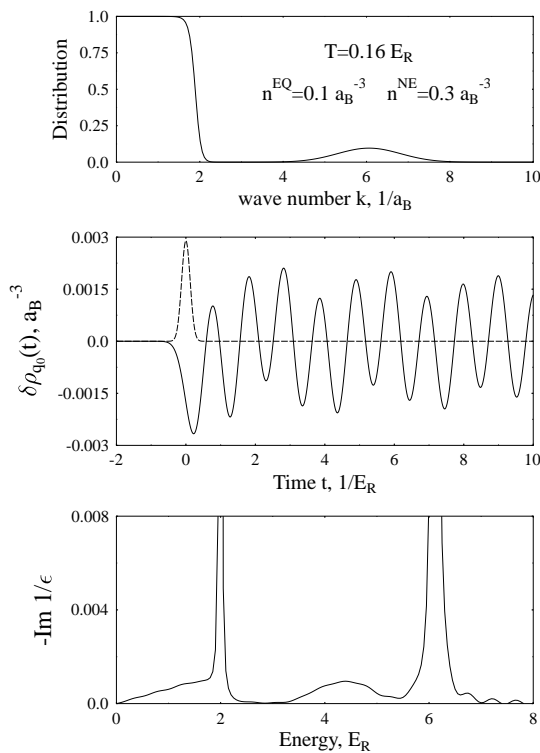, width=7.8cm}}
\vspace{-1cm}
	\caption{\small{Quantum kinetic results including RPA and
exchange for a strongly degenerate
electron gas in an optically excited n-doped semiconductor. {\bf upper fig.}: 
momentum distribution, with equilibrium (density $n^{EQ}$, temperature $T$) and
nonequilibrium electrons ($n^{NE}$). {\bf Middle fig.}: time-dependent total density
fluctuation (full line) excited by a monochromatic external field
(dashed line). {\bf Lower fig}:
structure factor with an optical 
(high frequency) and a weakly damped acoustic plasmon which emerges out of the
pair continuum.}}
	\label{fig2}
\end{floatingfigure}
\noindent
for very complex correlation effects by using very simple (compared e.g. to
the Bethe Salpeter theory) selfenergies in the collision integrals, and it
automatically yields all necessary vertex corrections. The numerical result in
Fig.1.b used the static Born approximation, thereby fulfulling the f-sum rule
to better than $0.1\%$.

To extend these calculations to plasmas with
{\em strong correlations and strong degeneracy}, one has two possibilities:
1. Generalize MD simulations to nonequilibrium quantum systems, e.g. \cite{qmd}.
2. Include strong correlation effects into the collision integrals of the
quantum kinetic equations. 
Alternatively,
one can use, instead of the selfenergies, pair correlation functions from
independent calculations \cite{qk-pimc}, such as path integral Monte Carlo
\cite{fb}. This should allow to obtain first-principle dynamic response »
functions in the near future.

\begin{acknowledgements}
This work is supported by the DFG
(Mercator-Programm, Schwerpunkt ``Laserfelder'' and ``Quantenkoh\"arenz in
Halbleitern'') and a grant for CPU time at the NIC J\"ulich.
\end{acknowledgements}


\end{document}